# EBG: A Lazy Functional Programming Language Implemented on the Java Virtual Machine

Tony Clark

**Abstract**

The Java programming language offers a number of features including: portability; graphics; networking. Java implements the object-oriented execution model in terms of classes, objects with state, message passing and inclusion polymorphism. This work aims to provide a mixed paradigm environment which offers the advantages of both object-oriented and functional programming. The functional paradigm is supported by a new language called EBG which compiles to the Java VM. The resulting environment can support applications which use both object-oriented and functional programming as appropriate.

## 1 Introduction

The programming language Java has become very popular by combining a number of features including portability, object-oriented programming, WWW compatibility, networking, graphics, and a growing collection of libraries. The language itself is reasonably small and offers a particular model of programming language execution based on classes, objects, message passing, and inclusion polymorphism (Cardelli & Wegner 1985).

Although the benefits of using the language are large, most notably its portability and ease of library construction, programmers are forced to use a particular style of programming, even when it does not suit all parts of the application. For example, operations over polymorphic lists are not readily supported by the object-oriented model since inclusion polymorphism is often incompatible with parametric polymorphism, Java uses *type casts* to recover the type of a list element. Another example occurs when programming in terms of lists whose elements are data items of loosely related data types, Java requires the use of *type tests* to determine the actual type of a data item.

Fortunately, the portability of Java arises from its use of a Virtual Machine (VM). This is a standard interface for executable code defined in terms of a collection of machine instructions. In principle, to take advantage of Java features it is not necessary to program in Java. So long as a program can be translated into Java VM instructions, it can offer Java-like advantages.

This paper describes research which aims to produce a mixed programming environment offering Java-like advantages. The environment provides a new language called EBG in addition to Java. EBG is a lazy, higher order functional programming language with a Hindley-Milner type system, modules, separate compilation, algebraic types, pattern matching, and an interface to Java based on the object-oriented model of program execution.

The resulting environment allows applications to be implemented as a mixture of functional and object-oriented programming with the aim being to allow control and data to pass (semi-) freely between the languages.

The essential feature of the implementation is to translate a functional program into an equivalent Java program using a one-to-one correspondence between functions and classes. Each execution of a function definition produces a new *closure*; correspondingly, the Java program instantiates the appropriate class producing an *object*. Since the Java VM does not directly support lexical scoping and nested classes (*class closures*), a process termed *class lifting* is performed on the Java program.

A new binary format is used to contain the result of transforming and compiling an EBG program. The default Java class loader is extended to recognise both the extended and basic formats allowing EBG and Java binary files to be loaded into the same machine. Finally, the Java reflective language features are exploited to allow EBG and Java programs to interact.

This paper is structured as follows. Section 2 provides example EBG program code and shows how the interface to Java programs is used. Section 3 describes how EBG code is translated to Java by defining interpreters for subsets of both languages and sketching a proof of consistency for the translation. The languages are called $\lambda$ and $\mu$Java respectively.

Section 4 describes how class lifting is performed which transforms a Java program containing nested classes into one in which classes occur only at the top-level. Section 5 describes how the EBG code is translated to Java VM code via an intermediate EBG VM language, the extensions to the class loader and the inter-language communication mechanisms. Finally, section 6 analyses the work, compares it with related work and outlines future plans.

A basic knowledge of Java, object-oriented programming and functional programming are assumed. The reader is directed to Garside & Mariani (1998), Venners (1998), Meyer (1988), Bird & Wadler (1988) and Field & Harrison (1988) for introductory material.

## 2 Example EBG Programs

### 2.1 Sieve of Eratosthenes

Figures 1 and 2 show a simple example of a mixed language application. Figure 1 is an EBG package called `Sieve` which implements a lazily generated list of prime numbers using a process called the *Sieve of Eratosthenes*, see Henderson (1980) for more details. The packages `list` and `command` provide definitions

```
import list, command;

integersFrom n = n:(integersFrom(n + 1));

sieve(n:ns) = n:sieve(remIf(\x. divisible x n) ns);

primes = sieve(integersFrom 2);

main =
  (new "TestSieve") $produces \Obj o.
  send o "printPrimes" []
```

Figure 1: Example EBG Code for Sieve of Eratosthenes

for list and Java interface operators respectively.

The package contains a collection of definitions. `integersFrom` is a function which generates an infinite list of numbers in sequence starting with `n`. `sieve` is a function which is applied to a list of numbers and removes those numbers which are multiples of numbers occurring earlier in the list. `primes` is a list of all prime numbers starting from 2.

The function `main` is an example of how imperative features are encoded in EBG. The command `new` takes a Java class name as an argument and instantiates the class. The infix operator `$produces` evaluates its left hand operand and supplies the value to its right hand operand. The command `send` is applied to an object, a method name and a list of arguments. The result is equivalent to the following Java statement:

```
o.printPrimes();
```

An EBG package roughly corresponds to a Java class where all of the top-level definitions are declared `static`. Any of the top-level symbols in an EBG package can be referenced by a Java program using the EBG package name as though it were a Java class name, for example `Sieve.primes`.

Figure 2 shows the source code for a Java class `TestSieve` which uses the EBG package `Sieve`. In addition, `TestSieve` uses a collection of static methods provided by `JavaInterface` which allow EBG values to be manipulated: `isList`; `isCons`; `head`; and `tail`.

Both EBG and Java source code compile, using the EBG compiler `ebgc` and the Java compiler `javac` respectively, to produce Java VM object code. Using a simple extension of the default Java class loader in addition to the package `java.lang.reflect`, both EBG and Java object code can be mixed into a running Java machine.

Execution of the system starts by loading the EBG `Sieve` package and starting to execute the commands in `main`. The first command creates an instance of the class `TestSieve` by dynamically loading the appropriate class file and instantiating the resulting class. The Java reflective interface is used to perform

```
public class TestSieve extends JavaInterface {
  public void printPrimes()
  {
    printNums(Sieve.primes);
  }
  public static void printNums(Value nums)
  {
    if(isList(nums))
      if(isCons(nums)) {
        System.out.println(head(nums));
        printNums(tail(nums));
      }
  }
}
```

Figure 2: Example Java Code Calling EBG Code

meta-level operations such as `send` which invokes a named method of an object. In this case when `printPrimes` is invoked control passes from EBG code to Java code.

The method `printPrimes` uses the EBG package as a class with a static attribute `primes` and calls `printNums` passing a lazily generated infinite sequence of prime numbers. The method `printNums` uses the methods `isCons`, `head` and `tail` to print out all of the elements of the list. The control flow of the program is shown in figure 3.

## 2.2 Environments

The evaluation of $\lambda$- and $\mu$Java-expression use environments to associate keys with values. In particular, free variables in an expression are bound in the current environment and $\mu$Java uses an environment to model the heap. Figure 4 shows the definition of an EBG package `Env` which implements environments. `env` is a parametric type with three data constructors. Type variables in EBG are sequences of `$` characters. `env` is parameterised with respect to the type of the keys and the type of the values.

An environment is either `Empty`, an association `Bind k v` between a key `k` and a value `v`, or the composition of two environments `Pair e1 e2`. Environment lookup is performed by:

`lookup key env default`

which returns the rightmost value associated with `key` in `env` or `default` if the environment does not contain the key. Environments may contain more than one entry for a key and shadowing occurs on the right. The function `mapEnv` is used to apply a function to all values in an environment.

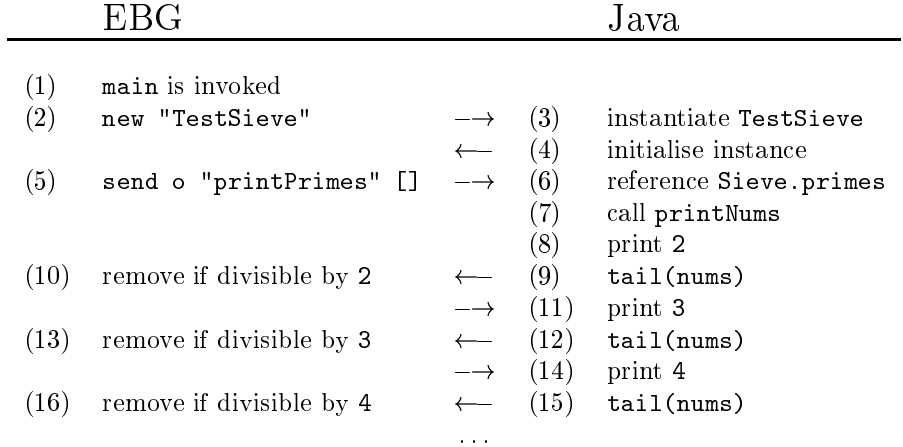

Figure 3: The control flow in Sieve of Eratosthenes

## 3 Compiling EBG

EBG is implemented by a translation to Java. The key issues of the translation are function representation and function application. This section describes these issues by defining two toy languages and analysing the translation between them.

The first language, called $\lambda$, is a sub-language of EBG providing integers, single argument functions, variables and function application. Its operational semantics is defined by an interpreter `ebgEval` written in EBG. The second language, called $\mu$Java, is a sub-language of Java providing nested class definitions and simple methods. Its operational semantics is defined by an interpreter `javaEval` written in EBG.

Compilation of EBG is modelled using a translation `trans1` from $\lambda$-programs to $\mu$Java programs. The translation is shown to be consistent (*i.e.* preserve the meaning of $\lambda$-programs) by defining a translation `trans2` from $\mu$Java values to $\lambda$-values such that the following diagram commutes (Sabry & Wadler 1997) :

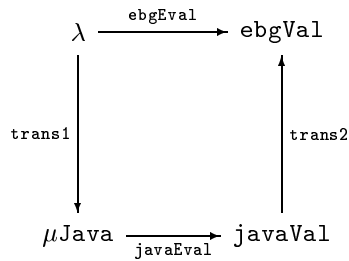

This section is structured as follows. Section 3.1 defines $\lambda$ and its operational semantics. Section 3.2 defined $\mu$Java and its operational semantics. Section 3.3 defines the translation `trans1` and sketches the proof of consistency.

```
type env $ $$ =
  Empty
| Bind $ $$
| Pair (env $ $$) (env $ $$);

lookup :: $ (env $ $$) $$ -> $$;

lookup key Empty default = default;

lookup key (Bind key' value) default =
  case key = key' of
   True -> value;
   False -> default
  end;

lookup key (Pair e1 e2) default =
  let value = lookup key e2 default
  in
    case value = default of
      True -> lookup key e1 default;
      False -> value
    end;

mapEnv :: ($ -> $$) (env $$$ $) -> env $$$ $$;

mapEnv fun Empty = Empty;

mapEnv fun (Bind key value) =
  Bind key (fun value);

mapEnv fun (Pair e1 e2) =
  Pair (mapEnv fun e1) (mapEnv fun e2)
```

Figure 4: Environment Structures

### 3.1 A $\lambda$-Calculus

EBG is a lazy functional programming language, therefore the operational semantics of $\lambda$ is based on a normal order reduction scheme (Hankin 1994, Plotkin 1975). The abstract syntax of $\lambda$ is defined as the type `ebg` in figure 5. The operational semantics is defined as a function `ebgEval` which is applied to a $\lambda$-expression and an environment associating variable names with thunks.

Evaluation of a $\lambda$-expression produces an integer, closure or an error. Note that well typed $\lambda$-expressions will not produce an error value. Figure 5 defines a type `ebgVal` for the results of program evaluation.

EBG uses normal order reduction which means that expressions are only evaluated if it is necessary to produce the final program outcome. This strategy is implemented by passing unevaluated expressions as function arguments. If

```
type ebg =
  EBGInt int
| EBGVar string
| Lambda string ebg
| Apply ebg ebg;

type ebgVal =
  EBGIntVal int
| Closure string (env string ebgVal) ebg
| Thunk (env string ebgVal) ebg
| EBGError;

ebgEval :: ebg (env string ebgVal) -> ebgVal;

ebgEval (EBGInt n) env = (EBGIntVal n);

ebgEval (Lambda arg body) env =
  Closure arg env body;

ebgEval (Apply e1 e2) env =
  let Closure arg env' body = ebgEval e1 env in
  let newEnv = Pair env' (Bind arg (Thunk env e2))
  in ebgEval body ;

ebgEval (EBGVar s) env =
  let Thunk env' body = lookup s env EBGError
  in ebgEval body env';
```

Figure 5: Definition of `ebgEval`

the value of the argument is ever required to construct the result of the function then the expression is *forced*.

Delayed evaluation of function arguments is implemented by constructing a *thunk*. A thunk associates a program expression with the current environment so that it can be evaluated at some later date. The current environment contains values for all of the free variables in the delayed expression.

As an example of normal order evaluation, consider the following $\lambda$-expressions:

```
W = Lambda "x" (Apply (EBGVar "x") (EBGVar "x"))

M = Apply (Lambda "x" (EBGInt 1)) (Apply W W)
```

An *eager* evaluation strategy fully evaluates the argument to a function before applying it. If M is evaluated eagerly the application of W to itself will not terminate. However, a *normal order* strategy will only evaluate an argument expression if it is required in the body of the function. In this case:

```
ebgEval M Empty = EBGIntVal 1
```

## 3.2 A $\mu$Java Calculus

In order to show how EBG is implemented in Java we show how $\lambda$-expressions are implemented in $\mu$Java which is a sub-language of Java containing just the required language features. In particular, the required features include:

- Anonymous and nested classes. Closures and thunks are implemented as objects. Java allows classes to be nested and implements static scoping rules which correspond to nested functions and thunks in $\lambda$-expressions. The syntax for instantiating anonymous Java classes (Flannagan 1997) is:

  ```
  new class-name () { class-body }
  ```

  which defines a sub-class of `class-name` and immediately instantiates it.

- Class instantiation. Each execution of a $\lambda$-function or application requires a new closure and thunk respectively. $\mu$Java represents closures and thunks as instances of classes.

- Message passing. Closure objects provide a method `apply` which is used to apply the closure to an argument. Thunk objects provide a method `force` which forces the thunk when its value is required.

- Object attributes. Lazy evaluation requires that $\lambda$-expressions are evaluated at most once. A thunk has a field `cache` which is used to cache the value of its delayed expression when it is forced.

- Self reference. To implement lazy evaluation a thunk checks whether it has forced its delayed expression. If not, it sends itself a message to force and then cache the result.

### 3.2.1 $\mu$Java Syntax and Values

Figure 6 defines the type `java` which is the abstract syntax of $\mu$Java. A $\mu$Java program is an environment of class definitions one of which must define a method called `main` with a single argument. Execution of a $\mu$Java program starts by calling the method `main` and evaluating its body with respect to the environment of top-level class definitions. The values produced by evaluating $\mu$Java programs are defined by `javaVal` in figure 7. The values are: classes; objects; integers; the null value; boolean values; and an error value.

A class definition contains variable references and, since definitions may be nested, a class captures the current context when it is created. The current context is an environment associating all variables freely referenced in the method bodies of the class with their current values.

Consider the class `Thunk` defined in figure 11. This is a typical abstract class since it defines a method `force` which calls a method `value` whose implementation is left to a sub-class of `Thunk`. The definition of `Thunk` is represented as an abstract syntax data value in EBG as follows:

```
       type java =
         Seq java java              ;;; sequenced commands.
       | JavaInt int                ;;; integer expression.
       | JavaVar string             ;;; variable reference.
       | NullClassDef               ;;; the ultimate super-class.
       | ClassDef                   ;;; a class definition:
           java                     ;;; the super-class.
           (list string)            ;;; the attributes.
           (env string methodDef)   ;;; the methods.
       | New java                   ;;; instantiation expression.
       | Send java string java      ;;; method invocation (1 arg).
       | Send0 java string          ;;; method invocation (0 args).
       | This                       ;;; self reference.
       | If java java java          ;;; conditional command.
       | Set string java            ;;; variable update.
       | Eql java java;             ;;; equality test.

       type methodDef =
         MethodDef string java      ;;; method (1 arg).
       | MethodDef0 java;           ;;; method (0 args).
```

Figure 6: $\mu$Java Syntax

```
ClassDef
  NullClassDef
  ["cache"]
  (Bind "force"
    (MethodDef0
      (If (Eql (JavaVar "cache") (JavaVar "null"))
          (Seq (Set "cache" (Send0 This "value"))
               (JavaVar "cache"))
          (JavaVar "cache"))))
```

The same definition may be evaluated more than once causing different contexts to be associated with the same class. Consider the class Closure defined in figure 11. Each sub-class of Closure must define a method called apply with a single argument. Nested classes are possible, for example the following corresponds to the curried function $M = \lambda x.\lambda y.xy$:

```
M = new Closure() {
      Value apply(Thunk x) {
        new Closure() {
          Value apply(Thunk y) {
            x.force().apply(y);
    }}}}
```

M contains the definition of two anonymous sub-classes of Closure. The outermost class is instantiated producing a Java object o which represents $M$. The outermost class contains no free variable references, however the innermost class

```
type javaVal =
  NullClass
| Class
    (env string int)
    javaVal
    (list string)
    (env string methodDef)
| JavaObjVal
    (env string method)
| JavaIntVal int
| Null
| JavaTrue
| JavaFalse
| JavaError string;

type method =
  Method
    string
    (env string int)
    javaVal
    java
| Method0 (env string int) javaVal java
| NoMethod;
```

Figure 7: $\mu$Java Values

contains a free reference to the variable x which is an argument of `apply` in the outermost class.

Each time o is sent an `apply` message, a new class is defined. In each case the class is associated with a different value for x. The following shows the class which is created as a result of `o.apply(t)`:

```
C = Class (Bind "x" t) Closure []
      (Bind "apply"
        (MethodDef "y"
          (Send
            (Send0 (JavaVar "x") "force")
            "apply"
            (JavaVar "y"))))
```

Notice that all $\mu$Java classes are associated with an environment, in this case `Bind "x" t`, which contains the values of variables which are freely referenced in the body of the class. For this reason we say that $\mu$Java supports *class closures*.

### 3.2.2 $\mu$Java Instantiation

Objects are environments associating method names with methods. A method has four components: an argument name; a captured context; an object; and a body. The context is an environment containing associations for all the freely referenced variables in the body of the method. The context is constructed when a class is instantiated by extending the class context with associations between the attribute names and their storage locations.

Each method contains an object which is used as the value of the pseudo-variable this. All methods in an object have the same object which is a cyclic reference to the object itself. Consider an object which is created when M from section 3.2.1 is evaluated. The object produced is referred to as o1 in the following $\mu$Java value:

```
o1 = JavaObj
  (Bind "apply"
    (Method "x" Empty o1
      (New (ClassDef Closure []
        (Bind "apply"
          (MethodDef "y"
            (Send
              (SendO (JavaVar "x") "force")
              "apply"
              (JavaVar "y"))))))))))
```

If the object o1 is sent an apply message with an argument t then the result is the class C in section 3.2.1. If C is instantiated the result is the following object o2 which captures the current context containing the value for x:

```
o2 = JavaObj (Bind "apply"
              (Method "y" (Bind "x" t) o2
                (Send
                  (SendO (JavaVar "x") "force")
                  "apply"
                  (JavaVar "y"))))
```

Class instantiation is performed by an EBG function instantiate expecting three arguments: the class to instantiate; a memory location used as the start of attribute storage; and an object to be used as the value of this. Instantiation produces three values: the new instance; an environment associating attribute names with storage locations; and the memory block used by the attributes. The value of this is found by a fixed point (Cook 1989, Clark 1994, 1996). If the value of instantiating the class c with respect to memory location l is o, a and h then instantiate satisfies the following equation:

```
(o,a,h) = instantiate c l o
```

Figure 8 shows the definition of the function instantiate. The process instantiates the super-class first and then merges the instance of the super-class with the extension attributes and methods to produce an instance of the sub-class.

### 3.2.3 Message Passing

Object-oriented program execution is performed using *message passing* which involves the lookup and invocation of an object's method. Message passing is performed using the function sendMessage expecting four arguments:

```
instantiate :: javaVal int javaVal -> (env string method,env string int,env int javaVal);

instantiate NullClass memoryLocation this = (Empty,Empty,Empty);

instantiate (Class env super atts meths) memoryLocation this =
  let (o1,a1,h1) = instantiate super memoryLocation this in
  let (a2,h2) = allocateAtts atts (memoryLocation + (usedMemory h1)) in
  let o2 = mapEnv (methodDefToMethod (Pair env (Pair a1 a2)) this) meths
  in (Pair o1 o2,Pair a1 a2,Pair h1 h2);
```

Figure 8: Definition of instantiate

```
sendMessage ::
  string
  (env string method)
  javaVal
  (env int javaVal) -> (javaVal,env int javaVal);

sendMessage message object value heap =
  case lookup message object NoMethod of
    Method name env this body ->
      let address = nextFreeMemoryLocation heap in
      let heap' = Pair heap (Bind address value);
          env' = Pair env (Bind name address)
      in javaEval body env' heap' this;
    NoMethod -> (JavaError message,heap)
  end;
```

Figure 9: Message passing in $\mu$Java

```
sendMessage message object value heap
```

where `message` is the name of the message, `object` is the target of the message, `value` is the value to be sent and `heap` is the current memory structure.

Messages are synchronous and the result of sending a message is a pair (value,heap') containing a data value and an updated memory.

Figure 9 shows the definition of message passing in $\mu$Java. The target is an environment and should associate the message name with a method. The method contains an argument name, an environment, an object and a program expression. The environment associates freely referenced variables in the body of the method with values. The environment is extended with the method argument and is used as the context for evaluating the method body.

### 3.2.4 $\mu$Java Evaluation

Evaluation of a $\mu$Java program `prog` is performed by:

```
javaEval prog env heap this
```

where `env` associates free variables in `prog` with memory addresses, `heap` associates memory addresses with java values, and `this` is an object whose method is currently being performed. Memory addresses are modelled as integers starting from 1. The $\mu$Java interpreter is shown in figure 10. It is defined by case analysis on the structure of the program. The interpreter 'threads' the heap through the program execution and produces a pair (value,heap') where value is a java value and, since the evaluation of `prog` can produce side effects, heap' is an updated heap.

## 3.3 Translation of $\lambda$-Terms to $\mu$Java

EBG is implemented by translating it to $\mu$Java. EBG closures are translated to instances of `Closure`, EBG thunks are translated to instances of `Thunk` and EBG integers are translated to $\mu$Java integers. This section defines the syntax translation from EBG programs to $\mu$Java programs and provides an overview of how values can be translated from one language to the other. These translations are then used to sketch the proof of consistency for the syntactic translation.

EBG values are integers or closures. The environments in closures associate variable names with thunks. EBG values are represented in $\mu$Java as instances of the classes defined in figure 11. The class `Value` is the super-class of all EBG values. The classes `IntVal`, `Closure` and `Thunk` define $\mu$Java representations of EBG integers, closures and thunks respectively.

The classes `Closure` and `Thunk` are abstract. EBG closures and thunks are defined as instances of sub-classes of these classes. Sub-classes of `Closure` must define a method called `apply` which is activated when the closure is applied. Sub-classes of `Thunk` must define a method called `value` which is activated when the thunk is forced for the first time. Once it is forced, an instance of `Thunk` uses the variable `cache` to retain the value.

```
javaEval ::
  java
  (env string int)
  (env int javaVal)
  javaVal -> (javaVal,env int javaVal)

javaEval (Seq j1 j2) env heap this =
  javaEval j2 env (2nd (javaEval j1 env heap this)) this;

javaEval (JavaInt n) env heap this = (JavaIntVal n,heap);

javaEval (JavaVar s) env heap this =
  (lookup (lookup s env 0) heap (JavaError "heap",heap);

javaEval NullClassDef env heap this = (NullClass,heap);

javaEval (ClassDef super atts meths) env heap this =
  let (class,heap') = javaEval super env heap this
  in (Class env class atts meths,heap');

javaEval (New j) env heap this =
  let (class,heap') = javaEval j env heap this in
  letrec (o,a,heap'') = instantiate class (loc heap') (JavaObjVal o)
  in (JavaObjVal o,Pair heap' heap'');

javaEval (Send exp message arg) env heap this =
  let (JavaObjVal o,heap') = javaEval exp env heap this in
  let (v,heap'') = javaEval arg env heap' this
  in sendMessage message o heap'';

javaEval (If exp1 exp2 exp3) env heap this =
  case javaEval exp1 env heap this of
    (JavaTrue,heap') -> javaEval exp2 env heap' this;
    (JavaFalse,heap') -> javaEval exp3 env heap' this;
  end;

javaEval (Set varName exp) env heap this =
  let (value,heap') = javaEval exp env heap this
  in (value,Pair heap' (Bind (lookup varName 0) value));

javaEval (Eql exp1 exp2) env heap this =
  let (value1,heap') = javaEval exp1 env heap this in
  let (value2,heap'') = javaEval exp2 env heap' this
  in case value1 = value2 of
       True -> (JavaTrue,heap'');
       False -> (JavaFalse,heap'')
     end;

javaEval This env heap this = (this,heap);
```

Figure 10: Definition of javaEval

```
abstract class Value {}

class IntVal extends Value {}

abstract class Closure extends Value {
  abstract Value apply(Thunk argument);
}

abstract class Thunk {
  private Value cache = null;
  public Value force() {
    if(cache == null)
      cache = value();
    return cache;
  }
  public abstract Value value();
}
```

Figure 11: EBG value classes

```
trans1 :: ebg -> java;

trans1(EBGInt n) = JavaInt n;

trans1(EBGVar s) = Send0 (JavaVar s) "force";

trans1(Lambda arg body) =
  New (ClassDef
        (JavaVar "Closure")
        []
        (Bind "apply" (MethodDef arg (trans1 body))));

trans1(Apply e1 e2) =
  Send (translate e1) "apply"
    (New (ClassDef
           (JavaVar "Thunk")
           []
           (Bind "value" (MethodDef0 (trans1 e2)))))
```

Figure 12: Definition of trans1

Translation of EBG programs to $\mu$Java programs is defined in figure 12. The translation of $\lambda$-functions and function application instantiate anonymous sub-classes of `Closure` and `Thunk` respectively. Function application is implemented using the method `apply` and thunks are forced using the method `force`.

Consider an EBG program `m` evaluated by `eval` with respect to an environment of thunks `e` producing an EBG value `v`. Given a translation `trans1` from environments of EBG thunks to environments of $\mu$Java objects, `m` and `e` can be translated and evaluated using `javaEval` to produce a $\mu$Java value `w` and a heap `h`. Given a translation `trans2` from $\mu$Java values and heaps to EBG values we must show that:

`ebgEval(m)(e)=trans2 o javaEval o trans1(m)(e)`

The proof is sketched as follows. EBG thunks are translated to produce instances of the appropriate sub-class of `Thunk`. Instances of `Thunk` and `Closure` are translated (relative to a heap) to EBG thunks and classes respectively. The proof of consistency proceeds by induction on the structure of the EBG program `m` and the environment `e`:

- If `m` is an integer then the proof follows by the definition of the interpreters and translations.

- If `m` is a variable then the proof follows by assuming that it holds for the body of the thunk bound to the variable in `e` and its environment.

- If `m` is a $\lambda$-function then the proof follows by assuming by induction that it holds for the body of the function and the environment `e`.

- If `m` is an application `Apply n1 n2` then we assume that the theorem holds for `n1`, `n2` with respect to `e` and also holds for the body of the resulting closure with respect to the extended closure environment.

## 4 Scope and Nested Classes

EBG is implemented in $\mu$Java using nested anonymous classes for both closures and thunks. Both $\mu$Java and EBG use lexical scoping rules for variable reference. Nested classes and lexical scoping rules are supported in $\mu$Java by class closures.

Although Java provides nested anonymous classes it does not implement class closures. In order to support lexical scoping it performs *class lifting* which is a process similar to *lambda lifting* (Field & Harrison 1988) in order to translate all class definitions to the top-level of the program. This section describes how EBG value classes are modified to take class lifting into account.

Class lifting is a Java program transformation whereby all classes are moved to the top-level. Lexical scoping is implemented by allocating space for variables in heap allocated activation frames. Consider the following $\lambda$-function:

$$M_1 = \lambda x.(\lambda y.yx)(\lambda z.xz)$$

```
class M1 extends Closure {
  Value apply(Thunk x) {
    frame = new Frame(x,frame);
    return new M2(frame).apply(new M3(frame));
  }
}

class M2 extends Closure {
  Value apply(Thunk y) {
    return y.force().apply(frame.local(0));
  }
}

class M3 extends Closure {
  Value apply(Thunk z) {
    return frame.local(0).force().apply(z);
  }
}
```

Figure 13: An example of class lifting

The result of $\lambda$-lifting $M$ is as follows:

$$M_1 = \lambda x.(M_2 x)(M_3 x)$$

$$M_2 = \lambda x.\lambda y.yx$$

$$M_3 = \lambda x.\lambda z.xz$$

The process of $\lambda$-lifting produces an equivalent program in which all nested functions have been moved to the top-level and extra parameters are added for their freely referenced variables.

Class lifting has the same effect as $\lambda$-lifting except that nested classes are moved to the top-level and variables are referenced via heap allocated frames. Figure 13 shows the result of translating $M_1$ to $\mu$Java and then performing class lifting. Note that the code in figure 13 has been simplified by omitting the creation of thunks. Section 5.2 describes the complete translation.

Class lifting is performed using the following algorithm. Let $P$ be a $\mu$Java program resulting from trans1. $P$ is a collection of class definitions indexed by their names. If $P$ contains no nested classes then stop. Otherwise a definition $d$ contains a nested class definition $c$. Depending on whether $c$ is a sub-class of Closure or Thunk, it may reference a single bound variable $v$ of $d$. Let $d'$ be $d$ with $c$ replaced by:

```
new k(new Frame(v,frame))
```

where $k$ is a new class name. Let $c'$ be $c$ with all references to $v$ replaced by:

```
abstract class Closure extends Value {
  private Frame frame;
  public Closure(Frame frame) {
    this.frame = frame;
  }
  public abstract Value apply(Thunk t);
}

abstract class Thunk {
  private Frame frame;
  private Value cache = null;
  public Thunk(Frame frame) {
    this.frame = frame;
  }
  public Value force() {
    if(cache == null)
      cache = value();
    return cache;
  }
  public abstract Value value();
}
```

Figure 14: Value class using frames

`frame.local(0)`

If $c$ references $v$ then all other expressions of the form `frame.local(n)` replaced with `frame.local(n+1)` The class $d$ is replaced with $d'$ in $P$ and $c'$ is added. This process is repeated until it terminates with no nested class definitions.

EBG value classes (initially defined in figure 11) are extended to support class lifting. Both `Closure` and `Thunk` are extended with an attribute `frame` whose value is supplied when an instance is created. The extended classes are shown in figure 14. `Frame` implements a linked list of values. The method `local` is used to index the list elements. The initial element in a frame is at position 0. New frames extend existing frames by adding a new element at the start of the list.

## 5 Implementation Issues

The semantics of EBG programs and their implementation in Java is defined by a consistent translation `trans1` in section 3.3. EBG is implemented by translating programs directly to Java VM code without generating any intermediate Java source code. The machine loader can freely mix Java and EBG object code and the reflective features of the Java machine permit Java and EBG code to interact. This section describes the implementation issues relating to the EBG environment.

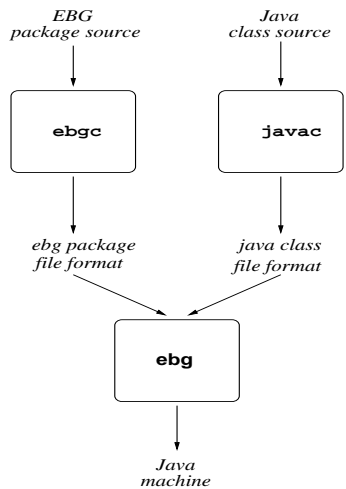

Figure 15: Mixing EBG and Java progrm code

## 5.1 The Class Loader

Java programs are executed by starting a Java machine and loading Java object files using a *class loader*. A class loader, running on the machine, is an object of type `ClassLoader` which is responsible for reading object files and linking Java VM code into the current running Java machine.

EBG defines a sub-class of `ClassLoader` called `ebg` which understands the format of both Java and EBG object files. The process of loading both EBG and Java into a running machine is shown in figure 15.

Compilation of a Java source file using `javac` produces an object file containing a binary representation in a *class file format*. There are entries in the binary file for all class components including fields, methods and static entries.

Compilation of an EBG source file using `ebgc` produces a file containing a binary representation in a *package file format*. The package file contains class file format entries for all the Java classes resulting from class lifting. In addition there is a distinguished class in each package which contains `static` fields for each top-level package definition. The value of each field is of type `Thunk` and both EBG and Java programs may reference any top-level EBG package definitions as static class fields. An EBG object package is an instance of the class `Package`:

```
public class Package implements Serializable
{
  public Vector importNames;
  public Hashtable classes;

  // Package methods ...
```

}

where `importNames` is a vector of imported package names and `classes` is a collection of associations between class names and arrays of bytes in Java class file format.

```
public class ebg extends ClassLoader
{
  private Hashtable classBytes = new Hashtable();
  private Hashtable loadedClasses = new Hashtable();
  private Vector importedPackages = new Vector();

  private void getClassBytes(String fName)
  {
    Package p = objStream(fName).readObject();
    Enumeration classNames = p.classNames();
    while(classNames.hasMoreElements()) {
      String cName = classNames.nextElement();
      classBytes.put(cName,p.classes.get(cName));
    }
    addElements(p.importNames,importedPackages);
  }

  private Class loadClass(String cName)
  {
    Class c;
    if(!loadedClasses.containsKey(cName))
      if(classBytes.containsKey(cName))
        c = defineClass(cName,classBytes.get(cName))
      else if(importedPackages.containsKey(cName)) {
        getClassBytes(cName);
        importedPackages.removeElement(cName);
        c = loadClass(cName);
      } else c = loadJavaClass(cName);
    else c = loadedClasses.get(cName);
    loadedClasses.put(cName,true);
    return c;
  }
}
```

Figure 16: The `ebg` Class Loader

A Java class is defined by a class loader by supplying the method `defineClass` with the name of the class and an array of bytes in class file format. Figure 16 shows the implementation of the EBG package loader `ebg`.

The EBG package loader uses three tables. The table `loadedClasses` is used to record when a class is loaded and defined. Once loaded and defined a class must not be re-defined. The table `classBytes` is used to hold the class file format byte codes of classes when EBG packages are loaded. The classes contained in an EBG package are defined on demand. Finally, the table `importedPackages` holds the names of packages which are imported but not yet loaded.

Once compiled, an EBG package is loaded using the extended class loader `ebg`. A package is loaded using the method `loadClass` which returns the Java class containing the EBG top-level definitions as static fields. The method `loadClass` uses the package loader tables to cache classes. Once a package is loaded, subsequent calls to `loadClass` will not need to re-load the package for different component classes.

## 5.2 Producing Java VM Code

EBG programs are compiled to Java VM code via an intermediate EBG VM language. The intermediate language allows the low level implementation to be changed without affecting the upper levels of the compilation process.

This section gives an overview of the EBG VM and the compilation process. In order to show the key features of the compilation three toy languages are used. EBG is modelled using the language $\lambda$ whose semantics is defined in section 3.1. EBG is compiled using an EBG function `compile` to produce EBG VM instructions implemented as an EBG data type `ebgInstr`. Translation to Java VM and class lifting is performed using an EBG function `trans3`. Given the semantics of Java VM, `javaVMEval`, the following diagram commutes:

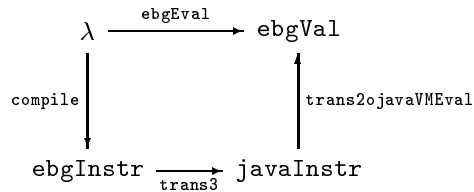

The EBG VM is a stack machine where the stack contains function activation frames. Each frame contains a code pointer to the current VM instruction, a pointer to the previous stack frame and the address of the current local variable frame. The machine instructions are defined as the type `ebgInstr` in figure 17. Compilation of an EBG program produces a sequence of EBG VM instructions. The compiler is defined in figure 17. A program is compiled as follows:

```
compile prog vars globals
```

where `prog` is an EBG program, `vars` is a list of variable names which occur freely in `prog`, and `globals` is an environment associating top-level variable names with the name of their defining package.

The Java VM is stack based. Each stack frame contains an object which is currently handling a message, a collection of locals, a pointer to the current VM instruction and a pointer to the previous stack frame. The object is always the value of local 0 and provides a collection of field values. In addition, the machine also contains a collection of classes which may be instantiated and whose static fields can be referenced.

```
type ebgInstr =
  PushInt int
| Local int
| Global string string
| PushLambda (list ebgInstr)
| App
| Force
| Delay (list ebgInstr);

compile ::
  ebg
  (list string)
  (env string string) -> (list ebgInstr);

compile(EBGInt n) vars globals =
  [ PushInt n ];

compile(EBGVar s) vars globals =
  case lookup s globals "" of
    "" -> [ Local (pos s vars), Force ];
    package -> [ Global package name, Force ]
  end;

compile(Lambda arg body) vars globals =
  [ PushLambda (compile body (arg:vars) globals) ];

compile(Apply exp1 exp2) vars globals =
  let
    instrs1 = compile exp1 vars;
    instrs2 = [ Delay (compile exp2 vars globals) ];
    instrs3 = [ App ]
  in instrs1 ++ instrs2 ++ instrs3
```

Figure 17: EBG Compilation

When $\lambda$ executes on the Java VM, the value of local 0 is always an instance of a sub-class of Closure or Thunk. The value of local 1 is always the current local frame.

Figure 18 shows an EBG type javaInstr whose values represent the Java machine instructions used to implement $\lambda$. The instructions are briefly explained as follows:

```
type javaInstr =
  VMNew int
| Aload0
| Aload1
| Astore1
| Bipush int
| GetStatic string string string
| Return
| InvokeVirtual string
| GetField string
| Dup
| InvokeSpecial string;

type VMClass =
  VMClosure int (list javaInstr)
| VMThunk int (list javaInstr);
```

Figure 18: Java VM Instructions

| | |
|---|---|
| VMNew n | instantiate the class named n |
| Aload0 | push the current object onto the stack |
| Aload1 | push the current local frame onto the stack |
| Astore1 | set the current local frame from the head of the stack |
| Bipush n | push the integer n into the stack |
| Return | return the value at the top of the stack from the current method call |
| InvokeVirtual m | call the method m where the target is on the stack below the arguments |
| GetField f | push the value of field f |
| Dup | duplicate the head of the stack |
| InvokeSpecial m | initialise the object at the head of the stack |

Translation of EBG VM instructions and class lifting is performed by the EBG function `trans3` defined in figure 19. A translation is:

`trans3 instr classes`

where `instr` is an EBG VM instruction and `classes` is a list of sub-classes of both `Closure` and `Thunk`. The elements of `classes` are produced by class lifting and are represented as values of type `VMClass` defined in figure 18. The names of these classes are modelled as integers in the translation. Translation produces a pair:

```
trans3 ::
  ebgInstr
  (list VMClass) -> (javaInstr,list VMClass);

trans3(PushInt n) classes = ([ Bipush n ],classes);

trans3(Local n) classes =
  ([ Aload1,
     Bipush n,
     InvokeVirtual "local(I)LValue;" ],
   classes);

trans3(Global package name) classes =
  ([GetStatic package name "LThunk;")],classes);

trans3(PushLambda instrs) classes =
  letrec
    name = length classes;
    c = VMClosure name (is ++ [Return]);
    (is,classes') = maptrans3 instrs (c:classes)
  in ([VMNew name,
       Dup,
       Aload1,
       InvokeSpecial "<init>(LFrame;)V"],
      classes');

trans3 App classes =
  ([InvokeVirtual "apply(LThunk;)LValue;"],classes);

trans3 Force classes =
  ([InvokeVirtual "force()LValue;"],classes);

trans3(Delay instrs) classes =
  letrec
    name = length classes;
    g = [ GetField "frame", Astore1 ];
    t = VMThunk name (g ++ is ++ [Return]);
    (is,classes') = maptrans3 instrs (t:classes)
  in ([VMNew name,
       Dup,
       Aload1,
       InvokeSpecial "<init>(LFrame;)V"],
      classes');
```

Figure 19: Translation of EBG VM to Java VM

```
[PushLambda
   [PushLambda
      [Local(1),
       Force,
       Delay
          [Local(2),
           Force],
       App],
    Delay
       [PushLambda
          [Local(2),
           Force,
           Delay
              [Local(1),
               Force],
           App]],
    App]]
```

Figure 20: EBG VM instructions for $M_1$

(instrs,classes')

where `instrs` is a list of Java VM instructions and `classes'` is an extended list of sub-classes. Figure 19 shows that the translation process macro-expands the EBG VM instructions and lifts classes each time a `PushLambda` or a `Delay` instruction is encountered.

Consider the $\lambda$-expression $M_1$ which is defined in section 4. Figure 20 shows the result of representing $M_1$ as a value of type `ebg` and then using `compile` to produce EBG VM instructions.

Figure 21 shows the classes produced by translating the EBG VM instructions to Java classes using `trans3`. The sub-classes of `Closure` labelled 0, 1 and 4 correspond to the functions $M_1$, $M_2$ and $M_3$ respectively. The sub-classes of `Thunk` labelled 2, 3 and 5 are used to delay the evaluation of function arguments.

## 5.3 Inter-language Communication

The EBG environment allows communication between EBG and Java code within the same Java machine. Communication occurs through the Java library `java.lang.reflect` which allows Java programs to manipulate and change themselves during program execution.

EBG packages are implemented as Java classes where the top-level definitions are encoded as static fields of type `Thunk`. When `ebg` loads the first EBG package it searches for the value of the field `main` and forces its value:

```
Field mainField = mainClass.getField("main");
Thunk mainThunk = (Thunk)mainField.get(null);
Class thunkClass = (Class)loadedClasses.get("Thunk");
Method force = thunkClass.getMethod("force");
```

```
  [VMNew(0),Dup,Aload1,
   InvokeSpecial(<init>(LFrame;)V)]

VMThunk 5
  [GetField(frame),Astore1,
   Aload1,Bipush(1),
   InvokeVirtual(local(I)LValue;),
   InvokeVirtual(force()LValue;),
   Return]

VMClosure 4
  [Aload1,
   Bipush(2),
   InvokeVirtual(local(I)LValue;),
   InvokeVirtual(force()LValue;),
   VMNew(5),
   Dup,Aload1,
   InvokeSpecial(<init>(LFrame;)V),
   InvokeVirtual(apply(LThunk;)LValue;),
   Return]

VMThunk 3
  [GetField(frame),Astore1,
   VMNew(4),
   Dup,Aload1,
   InvokeSpecial(<init>(LFrame;)V),
   Return]

VMThunk 2
  [GetField(frame),Astore1,
   Aload1,Bipush(2),
   InvokeVirtual(local(I)LValue;),
   InvokeVirtual(force()LValue;),
   Return]

VMClosure 1
  [Aload1,Bipush(1),
   InvokeVirtual(local(I)LValue;),
   InvokeVirtual(force()LValue;),
   VMNew(2),
   Dup,Aload1,
   InvokeSpecial(<init>(LFrame;)V),
   InvokeVirtual(apply(LThunk;)LValue;),
   Return]

VMClosure 0
  [VMNew(1),
   Dup,Aload1,
   InvokeSpecial(<init>(LFrame;)V),
   VMNew(3),
   Dup,Aload1,
   InvokeSpecial(<init>(LFrame;)V),
   InvokeVirtual(apply(LThunk;)LValue;),
   Return]
```

Figure 21: Java VM instructions for $M_1$

```
EBGsystem(force.invoke(mainThunk));
```

where `mainClass` is the class produced by `loadClass`, `mainThunk` is the value of `main` in `mainClass`, `force` is the method which forces thunk objects. The Java method `EBGsystem` is supplied with the result of forcing `mainThunk`.

`EBGsystem` is responsible for supplying the value of `main` with a sequence of Java VM responses to the sequence of requests which are generated. The model of EBG execution is shown below:

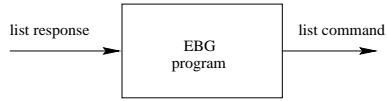

The commands produced by the definition of `main` in the package `Sieve` defined in figure 1 are `new` and `send`. The `new` command is handled by creating a new instance of the class `name` and adding it to the list of responses:

```
Class namedClass = Class.forName(name);
addResponse(namedClass.newInstance());
```

The `send` command is handled by finding the appropriate method called `name`, invoking the method with respect to the supplied `object` and `argVals` and then adding the return value to the list of responses:

```
Class objClass = object.getClass();
Method m = objClass.getMethod(name,argTypes);
Object[] args = new Object[]{argVals};
addResponse(m.invoke(object,args));
```

# 6 Conclusion

This work aims to provide a mixed paradigm programming environment which offers the advantages of functional programming (definition by cases, parametric polymorphism, lazy evaluation, higher-order functions, algebraic types) and the advantages of Java programming (object-oriented execution, inclusion polymorphism, portability, graphics, networking, multi-processing).

To achieve this aim, a new programming language called EBG has been designed and constructed. EBG offers many of the features of a modern functional programming language, compiles to the Java VM language and provides primitive features which allow the two languages to interact.

This paper has described the implementation of EBG in terms toy languages: $\lambda$; $\mu$Java; `ebgInstr`; and, `javaInstr`. These are sub-languages of the corresponding components of the real implementation whose features express the essential implementation characteristics.

In addition to those described in this paper EBG has a collection of standard functional programming features including: pattern matching in definitions and `case` expressions (Peyton Jones 1987) ; type checking and type inference (Cardelli 1984) ; and, named modules consisting of collection of type and value definitions which can be exported by the defining module and imported by other modules.

EBG functions have any number of arguments. The mechanism for maintaining local variables via instances of `Frame` is generalised to linked lists of heap allocated local frames where each frame has a number of entries corresponding to the function arguments.

EBG provides local variable binding using `case`, `let` and `letrec` expressions. In each case the compiler generates code which extends the current local frame with the appropriate number of values.

Compilation of EBG is very simple minded. This has the benefit that the interface between the two languages is clean; for example, closures and thunks can be passed freely between EBG and Java because they are implemented as Java objects.

In principle, closure-like and thunk-like objects can be created by Java as instances of sub-classes of `Closure` and `Thunk` then passed to EBG programs. This interface provides scope for experimenting with new types of 'function'; for example, functions can be created which connect to other Java machines over a network and which produce a stream of results.

The disadvantage of simple minded compilation is slow execution speeds for EBG programs. In addition, the Java VM code which is produced does not make efficient use of the Java VM stack, for example by passing function arguments via a stack frame rather than as part of instances of `Frame`.

EBG currently exists as a prototype implementation written in Java. The compiler uses the java compiler compiler `javacc`. The source code is currently about 20000 lines of Java code (around 3000 of which is automatically generated by `javacc`). EBG has been used to write a number of EBG libraries, some tutorial examples and the code in this paper.

The next phase of EBG work will address its compilation and the expansion of EBG VM instructions to Java VM instructions. In addition, functional programming research has produced a number of techniques for analysing and transforming programs in order to increase their speed and decrease their space usage. These techniques include: strictness analysis (Peyton Jones 1987) ; the STG machine (Peyton Jones 1992) ; and, deforestation (Wadler 1990) .

EBG is novel since it is a lazy functional programming language which compiles to the Java VM. Haskell evaluates lazily but does not compile to the Java VM. MLJ, developed by Persimmon IT, is a compiler for Standard ML which produces Java bytecodes. Standard ML is a higher order functional programming language with an eager evaluation strategy.

Kawa (Bothner 1998a 1998b) is an implementation of the lisp-derivative Scheme which compiles to the Java VM. Although Scheme employs an eager evaluation strategy, the translation of Kawa directly to the Java VM uses similar mechanisms to EBG. For example, Kawa implements Scheme procedures as instances of sub-classes of a Java abstract class `Procedure` which defines a collection of `apply` methods.

Pizza (Odersky & Wadler 1997) and more recently GJ (Brache *et al.* 1998)

are extensions of the Java language which aim to address the problem of parametric types. In the case of Pizza, Java is extended with parametric types (such as *list of anything*) which are incompatible with existing Java types (such as *list of* `Object`). GJ aims to extend Pizza so that both of these types have the same representation. Our approach differs in that we have provided parametric types in EBG which is a different language from Java but can be executed on the same machine. The lazy evaluation mechanism of EBG is not addressed by either Pizza or GJ.

Future plans for EBG include increasing the sophistication of its compilation and making the Java graphics, networking and multi-processing facilities available within a functional programming language.